\documentclass[prx, twocolumn,  superscriptaddress, longbibliography]{revtex4-2}
\usepackage{setspace}
\usepackage[table]{xcolor}
\usepackage{tabularx}
\usepackage[colorlinks, citecolor=blue, linkcolor=blue, urlcolor=blue, linktocpage]{hyperref}

\usepackage[utf8]{inputenc}
\usepackage{graphicx}
\usepackage{amsmath}
\usepackage[]{natbib}

\usepackage{graphicx}
\setcitestyle{numbers}

\usepackage{eurosym}
\usepackage{tikz}

\usepackage{enumitem}
\setitemize{noitemsep,topsep=0pt,parsep=0pt,partopsep=0pt,leftmargin=*}
\usepackage{amssymb}

\renewcommand*{\vec}[1]{\mathbf{#1}}

\renewcommand{\vec}[1]{\textbf{#1}}

\newcommand{\Fig}[1]{Fig.~\ref{#1}}

\usepackage{mdframed}
\usepackage{enumitem}

\begin{document}
\title{Quantum optimization with arbitrary connectivity using Rydberg atom arrays}

\author{Minh-Thi Nguyen}
\thanks{These authors contributed equally to this work.}
\affiliation{QuEra Computing Inc., 1284 Soldiers Field Road, Boston, MA, 02135, USA}

\author{Jin-Guo Liu}
\thanks{These authors contributed equally to this work.}
\affiliation{Department of Physics, Harvard University, Cambridge, Massachusetts 02138, USA}
\affiliation{QuEra Computing Inc., 1284 Soldiers Field Road, Boston, MA, 02135, USA}

\author{Jonathan Wurtz}
\affiliation{QuEra Computing Inc., 1284 Soldiers Field Road, Boston, MA, 02135, USA}

\author{Mikhail D. Lukin}
\affiliation{Department of Physics, Harvard University, Cambridge, Massachusetts 02138, USA}

\author{Sheng-Tao Wang}
\email{swang@quera.com}
\affiliation{QuEra Computing Inc., 1284 Soldiers Field Road, Boston, MA, 02135, USA}

\author{Hannes Pichler}
\email{hannes.pichler@uibk.ac.at}
\affiliation{Institute for Theoretical Physics, University of Innsbruck, Innsbruck A-6020, Austria}
\affiliation{Institute for Quantum Optics and Quantum Information, Austrian Academy of Sciences, Innsbruck A-6020, Austria}

\date{\today}
\begin{abstract}
Programmable quantum systems based on Rydberg atom arrays have recently been used for hardware-efficient tests of quantum optimization algorithms [Ebadi et al., Science, {\bf 376}, 1209 (2022)] with hundreds of qubits. In particular, the maximum independent set problem on 
 so-called unit-disk graphs, was shown to be efficiently encodable in such a quantum system. Here, we extend the classes of problems that can be efficiently encoded in Rydberg arrays by constructing explicit mappings from a wide class of problems to maximum weighted independent set problems on unit-disk graphs, with at most a quadratic overhead in the number of qubits. We analyze several examples, including: maximum weighted independent set on graphs with arbitrary connectivity, quadratic unconstrained binary optimization problems with arbitrary or restricted connectivity, and integer factorization. Numerical simulations on small system sizes indicate that the adiabatic time scale for solving the mapped problems is strongly correlated with that of the original problems.
Our work provides a blueprint for using Rydberg atom arrays to solve a wide range of combinatorial optimization problems with arbitrary connectivity, beyond the restrictions imposed by the hardware geometry.
\end{abstract}
\maketitle

\section{Introduction}\label{Sec:intro}
\begin{figure}[t!]
    \includegraphics[width=0.95\linewidth, trim={0cm 0.1cm 2cm 0cm},clip]{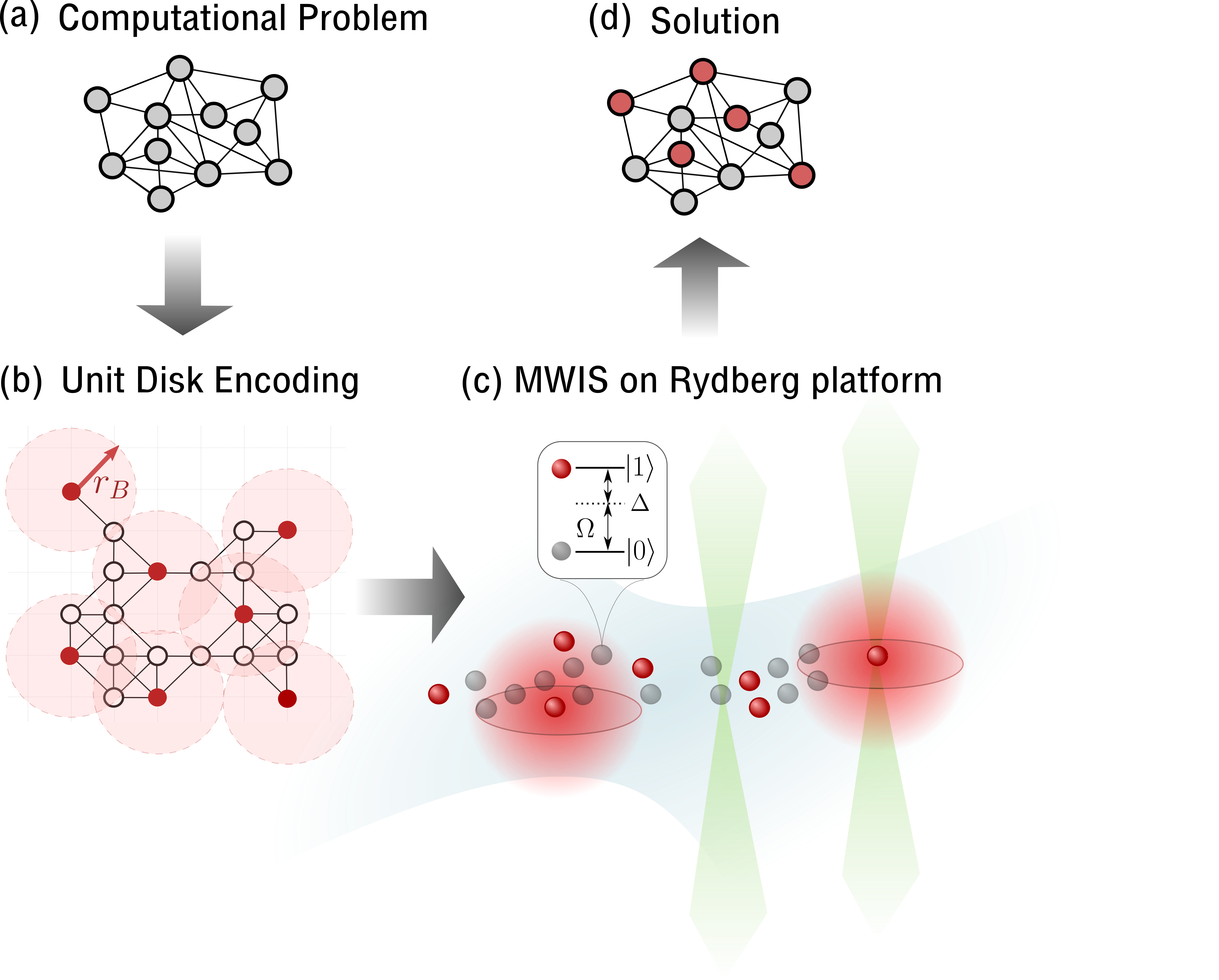}
    \caption{Procedure to solve a variety of optimization problems using programmable Rydberg atom arrays. The original computational problem (a) can be mapped onto a maximum-weight independent set (MWIS) problem on a unit-disk graph (UDG) in (b). (c) Physical platform, where each vertex in (b) represents an atom trapped by optical tweezers.  Each two-level atom can be coherently driven with Rabi frequency $\Omega$ and detuning $\Delta$, and the Rydberg blockade mechanism prevents two atoms from being simultaneously excited to state $| 1 \rangle$ if they are within a unit distance $r_B$. (d) The solution to the UDG-MWIS problem encodes the solution to the original problem.}
    \label{fig:platform_illustration}
\end{figure}

Quantum optimization algorithms aim to solve combinatorial optimization problems \cite{schrijver2003combinatorial, bernhard2008combinatorial} by utilizing controlled dynamics of quantum many-body systems. The key idea underlying this paradigm is to steer the dynamics of quantum systems such that their final states provide solutions to the optimization problem of interest. Such dynamics are often achieved either via the adiabatic principle in quantum annealing algorithms (QAA) \cite{farhi_quantum_2000, Farhi_2001, Kadowaki_1998, Das_2008, lidar_albash_review_2018}, or by employing more general, variational approaches, as exemplified by quantum approximate optimization algorithms (QAOA) \cite{farhi_quantum_2014}. A popular approach to design such quantum algorithms is to formulate the optimization problem in terms of a classical spin model~\cite{Lucas_Ising_Formulation} that can be implemented on special-purpose quantum hardware. 

An exciting possibility in this context is offered by Rydberg atom arrays \cite{Ebadi2022}. Owing to the Rydberg blockade mechanism \cite{jakschFastQuantumGates2000b, saffman2010, Ebadi2022}, these systems realize spin models that naturally encode a paradigmatic combinatorial optimization problem, namely the maximum independent set (MIS) problem on a special class of geometric graphs, called unit-disk graphs (UDG) \cite{Pichler2018MIS}. This allows a direct implementation of a variety of quantum optimization algorithms on this platform \cite{Pichler2018MIS,Ebadi2022,Kim2022Rydberg}. Remarkably, first experiments exploring this approach~\cite{Ebadi2022} observed a superlinear quantum speedup over optimized classical simulated annealing for finding exact solutions for some of the hardest accessible graphs. 
However, the restriction to unit disk graphs limits the applicability of this approach. Overcoming this limitation is one of the major challenges for exploring quantum optimization on a much wider range of optimization problems, including several problems of industrial relevance~\cite{Wurtz2022Industry}.

Approaches to extend the applicability of Rydberg atom arrays beyond UDGs have been recently explored in Refs.~\cite{Kim2022Rydberg, Byun2022}, but they are either limited to a specific class of graphs~\cite{Byun2022}, or require three-dimensional arrays~\cite{Kim2022Rydberg}, and both require bespoke encoding for each problem graph with unclear overhead for general graphs. Alternatively, other schemes that can map arbitrary non-local interactions into local ones have been proposed \cite{Lechner2015Quantum,Qiu2020Programmable}, but their implementations are experimentally even more demanding, requiring either 4-body interactions \cite{Lechner2015Quantum, Dlaska2022Quantum} or the use of tunable Ising interactions \cite{Qiu2020Programmable}.


\begin{figure*}[t!]
    \includegraphics[width=\linewidth, trim={0 17cm 3.5cm 0}]{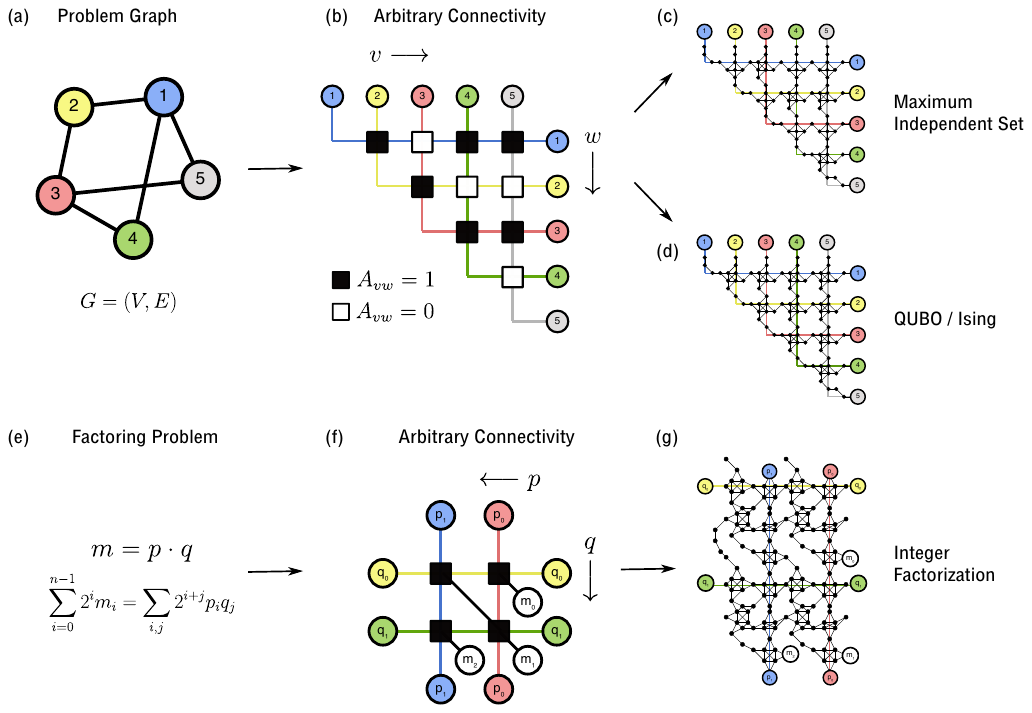}
    \caption{Architecture of UDG-MWIS mapping for three example problems.
    (a) Example non-UDG represented by $G = (V,E)$ for the MWIS and quadratic unconstrained binary optimization (QUBO) problems. (b) Crossing lattice used to construct UDG-MWIS mappings. Vertices in (a) are binary variables that can be represented effectively by lines to construct the lattice. Intersections in the lattice allow arbitrary connectivity between the variables, abstractly represented by squares. The lattice mimics an upper triangular adjacency matrix $A$, where for two vertices $\{v, w\} \in V$, $A_{vw} = 1$ if $(v, w) \in E $ and $A_{vw} = 0$ otherwise, represented abstractly by a filled and empty square here, respectively. (c) Final UDG-MWIS representation of the original MWIS problem on general graphs.
    (d) Final UDG-MWIS representation of the original QUBO/Ising problem. (e), (f), and (g) Similar encoding procedure for the integer factorization problem for the example $6 = 2 \times 3$, with the corresponding UDG-MWIS representation shown in (g).}
    \label{fig:mapping_overview}
\end{figure*}


In this paper, we introduce a new, systematic approach to encode optimization problems with arbitrary connectivity into Rydberg atom arrays. Our scheme requires only 2D atom trapping and the Rydberg blockade mechanism as main ingredients, both of which have been demonstrated already on current Rydberg atom array platforms with high fidelity~\cite{Ebadi2022} (see Fig.~\ref{fig:platform_illustration}). Importantly, our encoding is constructive and efficient as it incurs only a minimal, quadratic overhead in the number of qubits. We specifically discuss our approach on the paradigmatic optimization problems including maximum-weight independent set (MWIS) problems on arbitrary graphs, arbitrary quadratic unconstrained binary optimization (QUBO) or Ising problems. In addition, we apply our method to generic constraint satisfaction problems and show how integer factorization can be mapped to Rydberg atom arrangements. Finally, we perform  numerical simulations on small system sizes comparing the adiabatic time scale for original MWIS on non-UDG graphs to that of the mapped problem and observe strong correlations that suggest the encoding does not negatively impact the performance of quantum algorithms.

We note that MIS on UDGs is known to be NP-complete~\cite{clark1990}, so in principle any NP problem can be reduced to MIS on UDGs with a polynomial overhead. Specific, formal reduction sequences have, for example, been considered in Ref.~\cite{Ebadi2022}, but direct application of the prescribed reduction method requires at least $O(N^6)$ overhead. It is important for near-term implementation on quantum machines to find a low-overhead, explicit mapping, which is the main result of our work.



\section{Overview of Main Results}
In this section, we provide an overview of the main results of this work. The main ideas are summarized in Figs.~\ref{fig:platform_illustration} and \ref{fig:mapping_overview}. Given a computation problem, we map it to a UDG-MWIS problem (i.e., a MWIS problem on a UDG) using a novel encoding scheme. The resulting UDG-MWIS is the native problem for  Rydberg atom array platforms allowing a direct implementation of QAA or QAOA for its solution~\cite{Ebadi2022, Pichler2018MIS}. The solution for the UDG-MWIS obtained on the quantum device can then be mapped back to a solution for the original computation problem. The key result of this work is to provide a general framework for the low-overhead, efficient, and explicit mapping. 

The main idea underlying our encoding is summarized in Fig.~\ref{fig:mapping_overview} for three examples discussed in detail in this work: MWIS on general (non-unit-disk) graphs, QUBO/Ising problems with arbitrary connectivity, and integer factorization \footnote{We also show that circuit satisfiability problems can be mapped into UDG-MWIS and so all other NP problems can be mapped through circuit satisfiability if no better low-overhead mapping is found}. The general framework for mapping combinatorial optimization problems defined on graphs can be seen in Fig.~\ref{fig:mapping_overview}(a)-(d). First, the variables corresponding to vertices in the original graph can be encoded in one-dimensional chains of atoms using the ``copy gadget". These chains (represented by lines) are then arranged in the form of a crossing lattice shown in Fig.~\ref{fig:mapping_overview}(b), exhibiting exactly one crossing between each pair of lines. For each such crossing, we use additional gadgets---the ``crossing gadget" and the ``crossing-with-edge gadget"--- to encode the presence (and strength) or absence of an interaction for each pair of lines.  All these gadgets are carefully designed such that the resulting graph is a UDG (embedded on a square lattice), and the solution of the original problem is encoded in its MWIS. 
The mapping for the factoring problem follows a similar strategy (Fig.~\ref{fig:mapping_overview}(e)-(g)): We first encode the problem of finding the prime factors of a $N$-bit integer into an optimization problem; with the help of a crossing lattice and a properly designed factoring gadget, this optimization problem is then transformed to a UDG-MWIS problem. In all cases, the overhead in the number of qubits is at most $O(N^2)$, which is optimal for arbitrary connectivity~\cite{unweightedpaper}.

The manuscript is organized as follows. Sec.~\ref{sec:Background} introduces Rydberg atom arrays and explains the natural encoding of UDG-MWIS on the platform. Sec.~\ref{sec:encoding} outlines the basic encoding gadgets by first showing MWIS gadgets for simple constraint satisfaction problems. Then, some useful mapping gadgets are presented, including the ``copy gadget", the ``crossing gadget", and the ``crossing-with-edge gadget". Sec.~\ref{sec:arbitrary_Connectivity} uses the above gadgets and the idea of a crossing lattice to construct the explicit mapping to UDG-MWIS from three example applications: MWIS on general graphs, QUBO, and integer factorization. Additional gadgets and problem-specific details are presented. Sec.~\ref{sec:numerical_simulations} studies the performance of quantum algorithms before and after the mapping, focusing on the example of the MWIS problem from a general graph to UDG mapping. Finally, Sec.~\ref{sec:conclusions} concludes the paper and outlines some next steps and challenges. The appendix includes more details on strategies for overhead reduction and simplification in Appendix~\ref{sec:overhead_reduction}, a discussion on local defects and MWIS guarantees in Appendix~\ref{sec:normalization}, a more efficient mapping for problems with local connectivity in Appendix~\ref{sec:restricted_connectivity}, and more details on the construction of the factoring gadget in Appendix~\ref{sec:factoringAppendix}.

\section{Background}\label{sec:Background}
\subsection{Rydberg Atom Arrays}

This work is primarily motivated by recent advances in experiments with Rydberg atom arrays using neutral atoms in optical tweezers \cite{weimerRydbergQuantumSimulator2010a,  browaeys2020,kaufmanQuantumScienceOptical2021,Ebadi2021Quantum, Scholl2021Quantum,Byun2022,Graham_2022,madjarovHighfidelityEntanglementDetection2020,youngHalfminutescaleAtomicCoherence2020,singhDualElementTwoDimensionalAtom2022,singhDualElementTwoDimensionalAtom2022,zhangOpticalTweezerArray2022,steinertSpatiallyProgrammableSpin2022}. In these systems, atoms can be deterministically placed at programmable positions in two dimensions ~\cite{Ebadi2022, Ebadi2021Quantum, Scholl2021Quantum}. Each atom realizes a qubit with an internal ground state representing $| 0 \rangle$ and a highly excited, long-lived Rydberg state representing $| 1 \rangle$.  The atoms can be coherently manipulated with laser fields and interact pairwise via induced dipole-dipole interactions when two atoms are in the Rydberg state. Specifically, the laser induced quantum dynamics of this system can be described by the Hamiltonian
\begin{equation}
\label{eq:RydHamiltonian}
    H_{\text{Ryd}} = \sum_v \dfrac{\Omega_v}{2} \sigma^x_v - \sum_v \Delta_v n_v + \sum_{v < w}  V_{\text{Ryd}}(|\overrightarrow{\mathbf{r}_v} - \overrightarrow{\mathbf{r}_w}|)n_v n_w.
\end{equation}
Here, $\overrightarrow{\mathbf{r}_v}$ denotes the position of the atom labeled by $v$,  $\sigma^x_v = |1 \rangle _v \langle 0 | + |0 \rangle _v \langle 1 |$ coherently flips its internal state, and $n_v = |1 \rangle_v \langle 1 |$ counts if the atom is in the Rydberg state. The parameters $\Omega_v$ and $\Delta_v$ are the Rabi frequency and laser detuning for the $v$th atom. In experiments, the laser detuning can be controlled in a site-dependent way, for example, using local AC-Stark shifts \cite{Omran2019Generation}. 
The interaction potential $V_{\text{Ryd}}(|\overrightarrow{\mathbf{r}_v} - \overrightarrow{\mathbf{r}_w}|) = C_6/|\overrightarrow{\mathbf{r}_v} - \overrightarrow{\mathbf{r}_w}|^6$ 
leads to a strong (distance-dependent) energy penalty for configurations where two nearby atoms are simultaneously in the Rydberg state, giving rise to the so-called \textit{Rydberg blockade} mechanism \cite{jakschFastQuantumGates2000b, saffman2010,  Pichler2018MIS, Ebadi2022}. As a result, the low-energy states of the Hamiltonian do not contain states with pairs of  atoms that are both in the Rydberg state if they are within some characteristic distance, defined as the blockade radius $r_B$. This effect naturally imposes the independent set constraint on the ground state(s) of the Hamiltonian at $\Omega_v=0$, which, as discussed below, allows one to encode the MWIS of a corresponding UDG~\cite{Ebadi2022, Pichler2018MIS}. In this  classical limit ($\Omega_v=0$), it is convenient to  define the blockade radius $r_B$ via $V_{\rm Ryd}(r_B)=\max_v(\Delta_v)$, which is the convention we adopt throughout this work.  

\subsection{Unit Disk Graphs}
A unit disk graph is a graph $G=(V,E)$ with vertices $V$ and edges $E$ that can be embedded in the two-dimensional (2D) Euclidean plane such that two vertices are connected by an edge if and only if they are separated by a distance smaller than a unit radius. We are interested in unit disk graphs since they are in one-to-one correspondence with atom arrangements in 2D. Specifically, each atom represents a vertex, and we identify the blockade radius with the unit disk radius of the graph. In this way, the low-energy configurations of the atom array at $\Omega_v=0$ correspond to large independent sets of the unit disk graph \cite{Pichler2018MIS}. 

\subsection{Maximum Weight Independent Sets}
An independent set of a graph $G$ is the subset of vertices $S \subseteq V$, such that none of the vertices in $S$ are connected by an edge in $G$. The largest such independent set is called a maximum independent set. Note that in general the MIS may not be unique. The problem of finding a MIS is called the maximum independent set problem. The MIS problem can be generalized to the maximum weight independent set problem, where each vertex is assigned a weight $\delta_v>0$, and accordingly, a weight $W_S$ is assigned to each subset of vertices $S\subseteq V$ via $W_S=\sum_{v\in S}\delta_v$. The MWIS problem is to find an independent set with the largest weight.
It can be formulated as an energy minimization problem. For this, one can associate a binary variable $n_v \in \{ 0, 1 \}$ with each vertex $v \in V$. This allows us to identify a subset of vertices $S$ by a bitstring $\vec{n}=(n_1,n_2,\dots)$, via $S=\{v\in V|n_v=1\}$. We will frequently use this one-to-one correspondence between bitstings and subsets in the remainder of the paper. Using this representation, we can consider the cost function
\begin{equation}
H_\text{MWIS} = -\sum_{v \in V}\delta_v n_v + \sum_{(u, v) \in E} U_{uv} n_u n_v. \label{MIS_Hamiltonian}
\end{equation}
If $U_{uv} > \delta_w > 0$ for all $u, v, w$, the ground state configuration of $H_\text{MWIS}$ indeed corresponds to the MWIS. As in the unweighted case, the ground state can be degenerate, corresponding to multiple independent sets achieving the same maximum weight \footnote{Note that the particular scale of $U_{uv}$ does not change the low-energy properties (i.e., the energy of the states corresponding to independent sets) as long as $U_{uv} > \max_w \delta_w$.}.

If the graph $G$ is a UDG, we then refer to the corresponding MWIS problem as the UDG-MWIS problem. Importantly, for such UDG-MWIS problems, the Hamiltonian \eqref{MIS_Hamiltonian} coincides with the Rydberg atom array Hamiltonian \eqref{eq:RydHamiltonian} at $\Omega=0$, where each atom is placed at the respective location of the corresponding vertex, the blockade radius is identified with the unit disk radius (and the interaction tails beyond the blockade radius is neglected \cite{Pichler2018complexity, Ebadi2022}), and the weight of each vertex is identified with the local detuning $\Delta_v = \delta_v$~\cite{Ebadi2022, Pichler2018MIS}.  Hence, we aim in the following to encode a variety of problems of interests in UDG-MWIS.

\section{Encoding Gadgets}
\label{sec:encoding}

In this section, we construct a number of encoding gadgets, which are the basic tools used in this work to reformulate a variety of optimization problems as UDG-MWIS. To this end, we first encode solutions of a set of elementary constraint satisfaction problems as the solutions of a MWIS problem on properly constructed unit disk graphs.

\subsection{Constraint Satisfaction Problems as MWIS}\label{sec:CSFMWIS}

Consider a set of binary variables $\vec{n} = (n_1, n_2, \dots)$ with $n_i\in\{0,1\}$ and a set of constraints between them, denoted by $C$, that can be simultaneously satisfied by one or more  assignments. We represent this constraint satisfaction problem as a MWIS problem by constructing a weighted graph $G_C$, such that the constraint-satisfying assignments are in correspondence with the maximum weighted independent sets of $G_C$. More specifically, we say the MWIS problem on $G_C$ represents the constraint satisfaction problem $C$ if every MWIS of $G_C$ coincides with a satisfying assignment of $C$, and if every satisfying assignment of $C$ corresponds to at least one MWIS of $G_C$. Note that the number of vertices in $G_C$ can be larger than the number of variables in $C$, in which case we require the correspondence between the MWISs and the satisfying assignments only on the subset of vertices that correspond to the variables in $C$. Below, we illustrate this concept on several examples.

\subsubsection{Single Constraints}
\begin{figure}[t]
    \includegraphics[width=\linewidth, trim={0 21.8cm 12.2cm 0},clip]{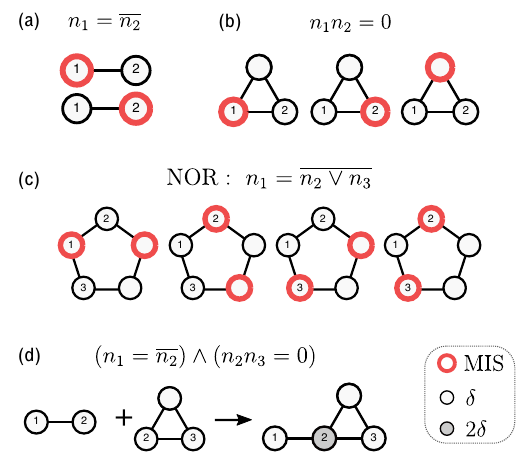}
    \caption{MWIS representation of some example constraints.
    Each bit is represented by a corresponding vertex in the MWIS problem graph. The weight of the vertices is indicated by its  interior color on a gray scale. For each example, the degenerate MWIS configurations are shown by identifying vertices in a MWIS with a red boundary. The MWISs correspond to the satisfying assignments to the corresponding constraint satisfaction problem.
    (a) MWIS representation of $n_1 = \overline{n_2}$. (b) MWIS representation of $n_1 n_2 = 0$, with the third, unlabeled vertex being an ancillary vertex. For each of the three MWIS states, the configuration on the relevant vertices 1 and 2 matches the corresponding satisfying assignment. (c) MWIS representation of the NOR constraint using 2 ancilla vertices. Note that only 4 of the 5 MWIS states are shown. Nevertheless, all 5 MWIS states correspond to the four satisfying assignments on the relevant vertices 1, 2 and 3.
    (d) A set of constraints $C=\{n_1=\overline{n_2}, n_2n_3=0\}$, where each is of the form given in (a) and (b). The corresponding MWIS problem is obtained by combining the two graphs corresponding to both constraints in $C$, resulting in the weighted graph on the right. Observe that vertex $2$, which appears in both constraints, has twice the weight of the other vertices. }
    \label{fig:CSP}
\end{figure}

We start with the simple example of two bits $\vec{n} = (n_1, n_2)$ with a single constraint
\begin{equation}\label{negation}
    n_1 = \overline{n_2},
\end{equation}
where $\overline{n_i}\equiv 1-n_i$ denotes the negation of $n_i$. This simple NOT constraint has two satisfying assignments, $\vec{n} = (1, 0)$ and $\vec{n} = (0, 1)$.

We represent this constraint satisfaction problem as a MWIS problem on a graph with two equally weighted vertices connected by an edge (\Fig{fig:CSP}(a)), whose cost function is simply $H_\text{MWIS} = -\delta(n_1 + n_2) + U n_1 n_2$, with $U > \delta > 0$.  This graph has two degenerate MWISs, $\vec{n} = (1, 0)$ and $\vec{n} = (0, 1)$, which are indeed in one-to-one correspondence to the two satisfying assignments of the constraint. 

Note, that for these two assignments, the cost function evaluates to $H_\text{MWIS} = -\delta$, while a violation of the constraint incurs a cost  $\delta>0$ (for  $\vec{n} = (0, 0)$) or a cost of $U - \delta>0$ (for  $\vec{n} = (1, 1)$), rendering them energetically unfavorable. For the remainder of the paper, we introduce the quantity $\delta_{\text{gap}} = \min( U - \delta, \delta) > 0$ as the minimum energy penalty for violation of a constraint. 

Next, we consider another simple two-variable constraint:
\begin{equation} \label{constraint}
    n_1 n_2 = 0.
\end{equation}
This constraint has three satisfying assignments $\vec{n} \in \{ (0,0), (1,0), (0, 1) \}$ and may be represented as a MWIS problem on a complete graph with three vertices with equal weights (\Fig{fig:CSP}(b)). The first two vertices, labelled $1$ and $2$, correspond to the two bits of interest, while the third vertex corresponds to an ancillary variable. 
The cost function associated with this MWIS problem is $H_\text{MWIS} = -\delta(n_1 + n_2 + n_3) + U(n_1 n_2 + n_2 n_3 + n_3 n_1)$, with the three degenerate solutions corresponding to the three satisfying assignments.  Importantly, each of the three MWIS states coincides with one of the three satisfying assignments on the two vertices of interest (see \Fig{fig:CSP}(b)). Again, a violation of the constraint incurs an energy cost of at least $\delta_{\text{gap}}$. 

We remark that in this manner one can construct the MWIS representation of all the basic operations in Boolean logic, by providing a gadget that is the MWIS representation of the NOR constraint in Fig.~\ref{fig:CSP}(c).

\subsubsection{Conjunction of Constraints}\label{conjuction_constraints}
Consider now a situation where $C$ consists of a set of multiple constraints that have to be satisfied simultaneously. For example, consider a conjunction of constraints involving three bits:
\begin{equation}
    (n_1 = \overline{n_2}) \ \land \  (n_2n_3 = 0). \label{ex_multipleconstraint}
\end{equation}
 which has three satisfying assignments, $(n_1,n_2,n_3)\in\{(1,0,0),(1,0,1),(0,1,0)\}$. To construct a corresponding MWIS representation, we first consider the two MWIS representations for the two involved constraints individually, which are given in Fig.~\ref{fig:CSP}(a) and (b), and then simply combine them by constructing the union of the individual graphs and add their weights (Fig.~\ref{fig:CSP}(d)). Equivalently, we add the two cost functions of the two individual constraints to obtain the MWIS cost function encoding of  Eq.~\eqref{ex_multipleconstraint}:
\begin{align}\label{eq:multipleconstraint_ham}
    \begin{split}
    H_\text{MWIS} = & -\delta(n_1 + 2n_2 + n_3 + n_4) + \\ 
    & U(n_1n_2 + n_2n_3 + n_3n_4 + n_4 n_2).
    \end{split}
\end{align}
It is easy to see that ground states of this cost function corresponds to the satisfying assignments in Eq.~\eqref{ex_multipleconstraint} on the vertices of interest, i.e. vertices 1, 2 and 3.

This example generalizes to the following important observation: Consider a set of constraints,  $C=\{C_1,C_2,\dots\}$, allowing for at least one satisfying assignment. Given the MWIS representations of each individual constraint $C_i$, we can construct a MWIS representation of $C$ by simply adding all the MWIS cost functions for all individual constraints in $C$.
The resulting cost function for $C$ indeed corresponds to a MWIS problem: its graph is simply the union of the individual graphs corresponding to the $C_i$s, with the corresponding weights added on the respective vertices \footnote{Note that, in general, this strategy could result in edges appearing in multiple constraints, producing inhomogeneous edge weights. However, because the satisfying assignments are independent sets, we can always homogenize the edge constraint by considering an equivalent problem of homogeneous edge weights by choosing a large enough $U\gg \delta$.}.

This is a powerful method that allows us to build MWIS representations of complicated constraints out of simpler ones.  We repeatedly make use of this technique in the following sections. The utility of this tool can already be illustrated by noting that the combination of the NOT and the NOR constraints (\Fig{fig:CSP}(a) and (c)) is universal. This immediately implies that we can encode any circuit satisfiability problem~\cite{Moore2011} into a MWIS problem with a constant overhead using this construction.

\subsection{Gadgets for Unit Disk Transformation} \label{gadgets}

While the representations introduced in the previous subsection allow the encoding of arbitrary constraint satisfaction problems into MWIS, additional gadgets are required for the specific task of transforming general graphs problems into UDG-MWIS problems, which can then been natively implemented using Rydberg atom arrays. Here, we introduce several particularly useful gadgets in this context.

\subsubsection{Copy Gadget and Effective Bits} \label{copy-gadget section}

\begin{figure}
\hspace{-0.1cm}
    \includegraphics[width=\linewidth, trim={0cm 17cm 12.5cm 0cm},clip]{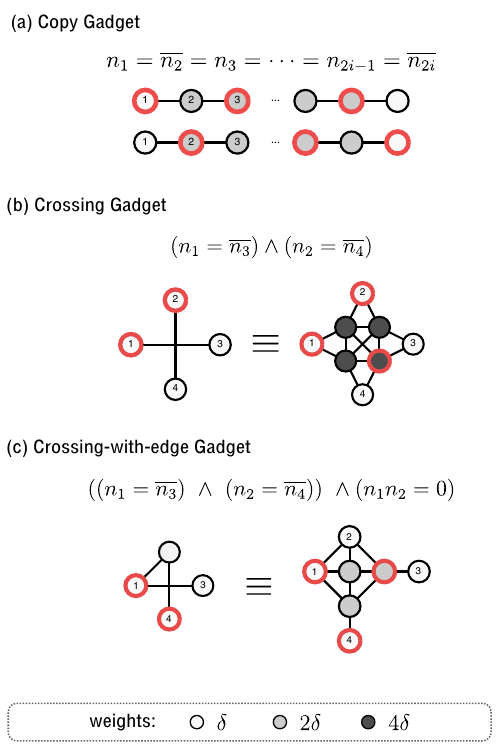}
  \caption{Important gadgets for formulating constraint satisfaction problems as UDG-MWIS. (a) Copy gadget. A 1D line graph encodes an effective bit. The two degenerate MWIS solutions are shown: the subset of odd-numbered vertices (top) and even-numbered vertices (bottom) represent the effective bit values $1$ and $0$, respectively. In this way, one can copy a single bit to any odd-numbered vertex. (b) Crossing gadget. The four degenerate MWIS solutions of the left graph coincide with the four MWIS solutions on the right graph on vertices $1,2,3,4$. One of these solutions is shown. Given a graph that contains a crossing, we can thus replace it with the right UDG, without changing the structure of the MWIS solution. (c) Crossing-with-edge gadget. Similar to (b), we can replace any subgraph of the type depicted on the left with the UDG on the right. One can check the MWIS solutions have one-to-one correspondence. The weights in (a)-(c) are encoded in grayscale according to the legend at the bottom of the figure.}
    \label{fig:udg-gadgets}
\end{figure}

By combining $N$ constraints of the form $n_m = \overline{n_{m+1}}$, we obtain a gadget, called the copy gadget: 
\begin{equation}\label{eq:copy}
n_1 = \overline{n_2} = n_3 = \overline{n_4} = \cdots .
\end{equation}
Here, the information of the bit $n_1$ is copied to all odd-index bits $n_3, n_5, \cdots$. As the name suggests, this copy gadget is useful in situations where the value of the bit $n_1$ is needed in several distant locations or in a location in conflict with the unit-disk requirement. Conceptually, copy gadgets ``stretch" the representation of a bit from a vertex (a point-like structure) to a one-dimensional line, while staying in the paradigm of unit-disk graphs. This technique is similar to other encoding approaches of using wires or chains of virtual vertices~\cite{Pichler2018complexity, Choi_2010, knysh2005, Qiu2020Programmable, Kim2022Rydberg}.

Using the techniques developed in Sec.~\ref{sec:CSFMWIS}, it is easy to construct the MWIS representation of the copy gadget in Eq.~\ref{eq:copy}. It consists of a one-dimensional graph with $N$ vertices and edges between neighboring vertices. All vertices have a weight $2\delta$, except for the two boundary vertices of the line, which have weights $\delta$ (see Fig~\ref{fig:udg-gadgets}(a)). Indeed, this weighted graph has two degenerate MWIS solutions: $\vec{n} = (1, 0, 1, 0, \cdots)$ and $\vec{n} = (0, 1, 0, 1,\cdots)$, corresponding to the two satisfying assignments of Eq.~\eqref{eq:copy}. We can thus use these two states to represent the effective binary variable with values 1 and 0 respectively (corresponding to the value of $n_1$). Note that the 1D line representation does not necessarily need to be drawn as a straight line when embedded in a 2D plane; it can bend and have kinks, as long as the resulting embedding satisfies the unit-disk criterion.  

Importantly, we can equip this effective bit with an effective weight $w$ by simply favoring one of the two staggered configurations with respect to the other. For instance, this can be achieved by adding an additional weight $w$ to any one of the equivalent vertices with an odd index in this gadget, e.g., the boundary vertex $n_1$. More generally, we can induce an effective weight $w$ by any vertex weight configuration as long as it satisfies $\sum_{m = 0}^i \delta_{2m + 1} = w + \sum_{m = 1}^i \delta_{2m}$ and $ 0 < \delta_m, w < U$. The latter inequalities ensure that the two staggered configurations remain the two lowest energy states, i.e., states with defect have a lower weight. 

\subsubsection{Crossing Gadget}

The copy gadget allows the effective representation of a binary variable as a 1D line on a UDG. When there are multiple such variables in a geometric representation, it can be extremely useful to allow two such lines to cross, without introducing any coupling between their corresponding effective degrees of freedom. However, such a crossing manifestly violates the unit-disk constraint. We solve this problem with the crossing gadget.

For this, consider the following set of constraints between four binary variables
\begin{equation}\label{eq:crossing}
    (n_1 = \overline{n_3}) \ \wedge \ (n_2 = \overline{n_4}).
\end{equation}
One way to represented this as a MWIS problem is to  identify each variable as an equally weighted vertex in a graph with edges $E = \{ (1, 3), (2, 4) \}$. Depending on the relative location of the  vertices in the 2D plane (which might be  fixed, e.g., due to additional constraints), these two edges might need to cross each other, violating the unit disk requirement. The crossing gadget is an alternative MWIS representation of the same pair of constraints \eqref{eq:crossing} that avoids this issue. This gadget is depicted in \Fig{fig:udg-gadgets}(b) and contains 4 ancillary binary variables (vertices). Note that the vertices representing the original variables are weighted equally with a weight $\delta$, while the four ancillary vertices have a weight $4\delta$ \footnote{Any weight larger than $2\delta$ for the ancillary vertices would work. The choice $4\delta$ is convenient, since it homogenized defect energies when crossing gadgets are combined with copy gadgets.}.
This graph is manifestly a UDG and realizes the desired relative geometrical distribution of the vertices 1, 2, 3 and 4. One can easily check that it has four-fold degenerate MWISs, which correspond to the four satisfying assignments on the four original vertices. 

\begin{figure}
    \includegraphics[width=1.0\linewidth]{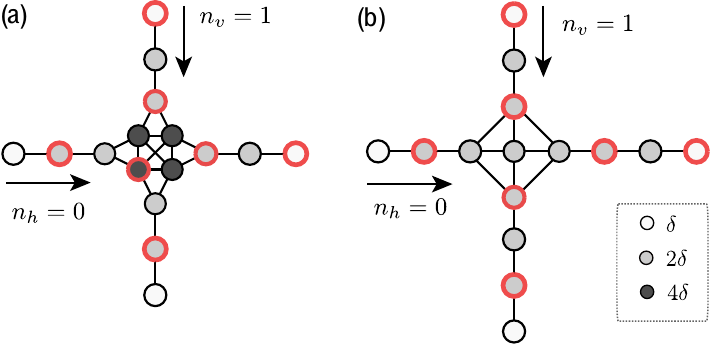}
  \caption{(a) Two decoupled effective degrees of freedom. By combining the copy gadget and the crossing gadget, we can form two 1D lines, here drawn horizontally and vertically. Both lines represent a binary variable, $n_h\in\{0,1\}$ and $n_v\in\{0,1\}$, respectively. The crossing gadget effectively decouples these degrees of freedom. Accordingly, this weighted graph has a 4-fold degenerate MWIS, corresponding to the 4 possible states of two binary variables. One of the MWISs is shown in red corresponding to $n_h=0$ and $n_v=1$. Note that each of the four MWISs contains exactly one of the 4 internal vertices of the crossing gadget, so these internal vertices encode the states of both effective degrees of freedom. (b) Two effective degrees of freedom, defined on a horizontal and a vertical line respectively, that satisfy an independence constraint $n_hn_v=0$, introduced by the crossing-with-edge gadget (Fig.~\ref{fig:udg-gadgets}(c)). Note that this graph has exactly 3 degenerate MWIS (out of which, one is shown in red), corresponding to the configurations $(n_h,n_v)\in\{(0,0),(0,1),(1,0)\}$.   }
    \label{fig:crossing_effective}
\end{figure}

In Fig.~\ref{fig:crossing_effective}(a), we illustrate how to combine a crossing gadget and the copy gadget to define two decoupled effective binary degrees of freedom living on two lines, a horizontal one and a vertical one. Specifically, in Fig.~\ref{fig:crossing_effective}(a), the two effective degrees of freedom are realized by the two staggered configurations of the horizontal and vertical line, and the crossing gadget decouples them; this structure is realized by extending each boundary vertex of the crossing gadget using the copy gadget. Following the recipe given in Sec.~\ref{conjuction_constraints}, one defines a vertex weight pattern with weights $4\delta$ on the interior vertices of the crossing gadget, weight  $2\delta$ on the exterior vertices of the crossing gadget and on all vertices of the lines, except for the boundary vertices of each line, which have a weight $\delta$.

By generalizing this example, we can see that the copy gadget allows us to represent binary variables as lines and the crossing gadget allows these lines to cross without introducing any interactions or constraints between their effective degrees of freedom, so we can arrange these effective 1D lines arbitrarily in 2D without worrying about crossings between them.

\subsubsection{Crossing-with-edge Gadget}
The crossing gadget is useful to decouple effective degrees of freedom defined on lines, even if the lines cross. In contrast, we now introduce a gadget that allows us to introduce a specific type of interactions between the effective degrees of freedom. Specifically, we are interested in a gadget that introduces the independence constraint $n_u n_v = 0$ between two effective variables, $n_u$ and $n_v$, when their corresponding lines cross. For this, we consider the situation where four binary variables must satisfy the constraints
\begin{equation}
    ((n_1 = \overline{n_3}) \ \wedge \ (n_2 = \overline{n_4})) \ \wedge \ (n_1 n_2 = 0), \label{eq:crossing_with_edge}
\end{equation}
where in this case, $n_u\equiv n_1$, $n_v\equiv n_2$. This  corresponds to the MWIS problem on the graph on the left of Fig.~\ref{fig:udg-gadgets}(c). In particular, we consider the situation where the vertices are geometrically positioned relative to each other in a way that requires a crossing; this case indeed occurs when $n_1, n_3$ and $n_2, n_4$ each belong to a line (created by a copy gadget) that corresponds to the effective binary variable associated with $n_1$ and $n_2$. This graph is, however, not a unit-disk graph, so we introduce the crossing-with-edge gadget shown on the right of Fig.~\ref{fig:udg-gadgets}(c). The resulting graph is manifestly a unit-disk graph with a three-fold degenerate MWIS solutions, corresponding to the three satisfying assignments of the crossing-with-edge constraint required in Eq.~\eqref{eq:crossing_with_edge}.

Analogous to the discussion of the crossing gadget, we can also combine the crossing-with-edge gadget with copy gadgets to obtain two crossing lines (drawn horizontally and vertically in Fig.~\ref{fig:crossing_effective}(b)) that host two effective binary degrees of freedom respectively, with an independence constraint between them. The resulting weight pattern is shown in Fig.~\ref{fig:crossing_effective}(b).

\section{Arbitrary Connectivity}
\label{sec:arbitrary_Connectivity}

Using the suite of encoding gadgets introduced in the previous section, we can now encode a variety of computational problems into UDG-MWIS, which can then be readily implemented on Rydberg atom arrays. Here, in this section, we discuss three example applications in detail: MWIS on graphs with arbitrary connectivity, QUBO problem, and the integer factorization problem. As we will see later, the resulting UDGs can be embedded on a square lattice with at most a quadratic overhead. The recipe involves two main steps: the first is to construct the so-called \textit{crossing lattice} using the copy gadget, and the second is to apply crossing replacements to encode arbitrary connectivity.

\begin{figure}[ht!]
    \includegraphics[width=\linewidth,  trim={0 18.6cm 12.8cm 0},clip]{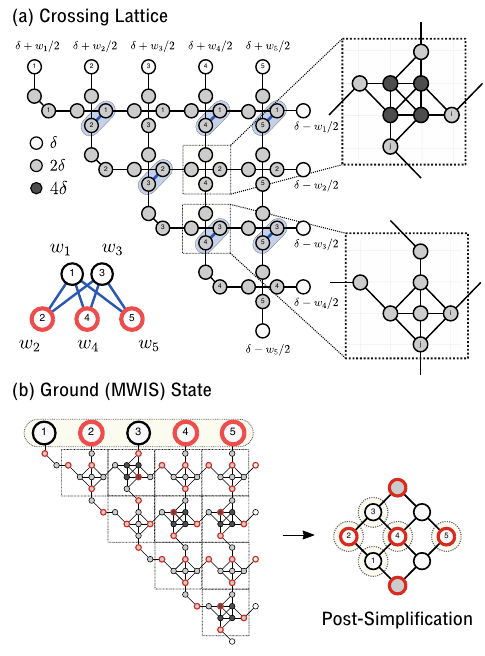}
\caption{Example encoding procedure for the MWIS problem on the $K_{2,3}$ (non-unit-disk) graph into UDG-MWIS. (a) Crossing lattice. Each vertex $v$ is represented by a line using the copy gadget, with odd-numbered vertices labeled. Each line is bent to form a triangular lattice, giving a crossing point between any two variables. If the two variables share an edge in the original graph, we draw an edge between their representative vertices on the lattice. A UDG is then obtained by replacing each crossing with a crossing or crossing-with-edge gadget. (b) UDG representation and the corresponding ground state solution. Each crossing in (a) is replaced with a unit cell containing at most 8 vertices, thus resulting in a final mapping with 92 vertices. The MWIS of the original graph ($\{2, 4, 5\}$) can be read out from the boundary vertices of the ground state of the mapped graph. This mapping can further be simplified (see Fig.~\ref{fig:simplificationK23}) to a mapping of only 9 vertices, where the vertices corresponding to the original degrees of freedom still encode the desired MWIS solution.}
    \label{fig:MWIS}
\end{figure}

\subsection{The Crossing Lattice}
\label{sec:crossing_lattice}
Consider an optimization problem for $N$ binary variables, such as the MWIS problem defined on an arbitrary graph $G = (V, E)$, where each vertex represents a binary degree of freedom.
To realize arbitrary connectivity, we first use the copy gadget to represent each vertex $v \in V$ by a 1D vertex line. The state of the binary variable associated with a vertex $v$ (0 or 1) can then be accessed at any odd-index vertex of the corresponding line (\Fig{fig:udg-gadgets}(a)). As detailed below, interactions between the effective degrees of freedom represented by these lines can be introduce at points where the lines cross. To achieve arbitrary connectivity, each line must thus cross every other line at least once. A simple layout achieving this is shown abstractly in \Fig{fig:mapping_overview}(b), where each line is drawn with a vertical and a horizontal segment, forming an upper triangular crossing lattice. In this way, a line (representing vertex $v$) crosses any other line (representing vertex $w$) exactly once. At these crossing points, we can then use the various crossing gadgets introduced in the previous section to induce interactions between $v$ and $w$ or to keep them decoupled. This is detailed concretely in Secs.~\ref{sec:MWIS_section} and \ref{sec:Ising} below. In addition to introducing interactions, these gadgets also turn the resulting graph explicitly into a UDG. 
Note that the resulting graph can be constructed by $N(N-1)/2$ ``tiles'', each containing 8 vertices for a tile formed by a crossing gadget, or 7 vertices for a tile formed by a crossing-with-edge gadget. Taking into account also the boundary vertices, we conclude that this construction leads to a UDG with at most $4N^2$ vertices, corresponding to the optimal quadratic overhead for arbitrary connectivity~\cite{unweightedpaper}.
We note that this particular choice of ``weaving" lines together may be sub-optimal, especially if the connectivity is sparse. In such a case, the total size of the lattice and thus the encoding overhead can be reduced by forming more sophisticated crossing lattices. More details of the simplification steps are included in Appendix~\ref{sec:overhead_reduction}.

\subsection{Maximum Weight Independent Set}
\label{sec:MWIS_section}

Building on the above recipe, we now detail how to map the MWIS problem on an arbitrary weighted graph $G=(V,E)$ with vertex weights $w_v$ ($v\in V$) to a UDG-MWIS problem. As shown in \Fig{fig:MWIS}(a), we first create a crossing lattice using the copy gadget.
At each crossing point between two lines (corresponding to vertices $u,v\in V$), we decouple the effective degrees of freedom using a crossing gadget if $(u,v)\notin E$, or induce an independence constraint for the effective degrees of freedom via a crossing-with-edge gadget if $(u,v)\in E$, as shown in \Fig{fig:MWIS}(b). We note that this results in a UDG that can be embedded on a square lattice (whose diagonal sets the unit disk radius). The weights on the vertices of this UDG are $2\delta$ on all vertices except the 4 interior vertices of the crossing gadget (which have weight $4\delta$) and the 2 boundary vertices of each line, whose weights are chosen to introduce the correct effective weights $w_v$ for the effective variable represented by each line. We chose a convention in which the first vertex along the line $v$ has a weight $\delta+w_v/2$, and the last vertex along this line has a weight $\delta-w_v/2$. 
Note that, to guarantee that the MWIS of the resulting UDG is indeed formed by the proper low energy configurations of each line, the weights have to satisfy $2w_v\leq \delta$. This can always be achieved by a proper normalization of the weights, or a suitable choice of $\delta$ (for more details see Appendix \ref{sec:normalization}). 

The MWIS of the mapped problem can be straightforwardly transformed back to a valid solution in the original problem. Indeed, since the state of the effective degrees of freedom associated with each line can be accessed at the first vertex of the line, the MWIS of the original problem is directly given by the configuration of the boundary vertices of the MWIS of the mapped problem. This solution readout is shown as the yellow ellipses of Fig.~\ref{fig:MWIS}(b).

\subsection{Quadratic Unconstrained Binary Optimization}\label{sec:Ising}

\begin{figure}[t!]
\hspace{-0.8cm}
    \centering
    \includegraphics[width=\linewidth,  trim={0 14.4cm 13.3cm 0},clip]{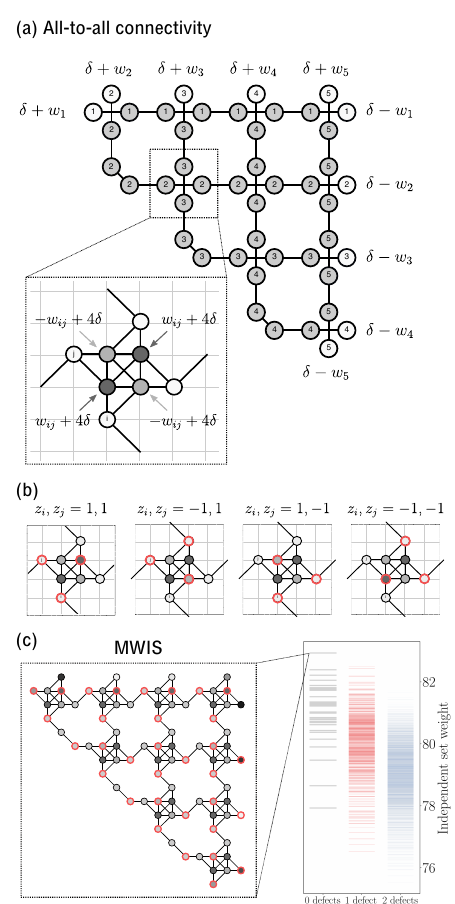}
     \caption{Example encoding procedure for the QUBO/Ising problem for a 5-bit system. (a) Crossing lattice.  Similar to the MWIS mapping, we can construct the UDG-MWIS representation of a generic QUBO problem by inserting a gadget at each crossing. The gadget has a similar structure as the crossing gadget used in the MWIS encoding, but the weights on the ancillary vertices are biased to induce quadratic interaction terms $w_{ij}$  between the effective degrees of freedom; see (b). (c) Example of an encoded $N=5$ QUBO problem, and the high-weight spectrum of the encoded cost function, illustrating that the MWIS is indeed in the 0-defect sector and thus encodes the solution of the QUBO problem. The QUBO solution $\{-1, +1, +1, +1, +1\}$ is encoded in the boundary of the graph. }
    \label{fig:weighted_crossing_gadget}
\end{figure}

QUBO is a paradigmatic NP-hard combinatorial optimization problem that has a wide range of applications. Generally, it seeks to find an input configuration that minimizes a quadratic polynomial function

\begin{equation}
    f(z) = \sum_{i<j}J_{ij}z_i z_j + \sum_i h_i z_i,
\end{equation}
where the domain of $f$ is binary bitstrings $z\in \{\pm 1\}^N$. QUBO is also called the Ising problem, where each bit can be represented by a spin $1/2$ degree of freedom, and the QUBO solutions correspond to the ground states of the Ising model.

To encode the QUBO problem in a UDG-MWIS, we again start with constructing the crossing lattice, with each line encoding one of the binary variables $z_i$ (Fig.~\ref{fig:weighted_crossing_gadget}(a)). For simplicity, we choose the number of vertices along each line to be even.
We then use the crossing gadget at each of the crossing points of the lattice, which decouples all the $N$ effective binary degrees of freedom. Recall that at this point in the construction all vertices in the resulting graph have a weight $2\delta$, except for the boundary vertices on each line, which have a weight $\delta$ and the 4 interior vertices of each crossing gadgets, which have a weight $4\delta$. The QUBO cost function is then imposed on the effective degrees of freedom by adjusting these weights as follows: Firstly, the weight of the two boundary vertices of line $i$ is adjusted to $\delta+ w_i$ for the first vertex and $\delta-w_i$ for the last one (see Fig.~\ref{fig:weighted_crossing_gadget}(a)). It is easy to see that for lines of even length this induces the linear term $h_i z_i$ for the effective degree of freedom $z_i$, with $w_i=h_i$ up to normalization (see Appendix~\ref{sec:normalization}). Secondly, the weights of the internal four vertices of the crossover gadget between the lines representing the bits $i$ and $j$ are adjusted to $4\delta\pm w_{ij}$ as depicted in the inset of Fig.~\ref{fig:weighted_crossing_gadget}(a), where $w_{ij}=J_{ij}$ up to normalization. To see that this induces the quadratic interaction term $J_{ij}z_iz_j$ between the two effective bits, recall that exactly one of the four ancillary vertices of the crossing gadget is part of the MWIS, and that this vertex is determined by the configuration of the effective degrees of freedom $z_i$ and $z_j$ (see Figs.~\ref{fig:crossing_effective}(a) and \ref{fig:weighted_crossing_gadget}(b)).  Similar to the MWIS encoding, the additional weights have to be appropriately normalized, such that the ground state of the cost Hamiltonian consists of configurations that correspond to valid (i.e., defect-free) configurations of the effective degrees of freedom. This is guaranteed by a normalization such that $\max_i(\sum_{j}|w_{ij}|,|w_i|)<\delta$. For more details on the normalization, see Appendix \ref{sec:normalization}.
Similar to the MWIS problem, the ground state of the QUBO problem can be directly inferred from the ground state of the resulting UDG-MWIS problem. An example is shown in Fig.~\ref{fig:weighted_crossing_gadget}(c) highlighting the MWIS for a random choice of $J_{ij}$ and $h_i$, and $N=5$. In the inset, we confirm that the MWIS state is indeed the one corresponding to the solution of the QUBO problem. We also show the weights of other independent sets, including those that do not correspond to valid configurations of the effective degrees of freedom, i.e., configurations that include defects. We note that the weights of the configurations in the zero-defect sector have a one-to-one correspondence with the spectra of the original QUBO problem. One can see that some states with defects have a higher total weight than some states that represent valid configurations of the effective variables (i.e., without defects), but, importantly, the MWIS is guaranteed to be in the zero-defect sector given proper normalization.

In summary, any QUBO problem on $N$ variables can be encoded in a UDG-MWIS problem with at most $4N^2+\mathcal{O}(N)$ vertices. For restricted connectivity, one may construct a lower-overhead crossing lattice; for details, see Appendix~\ref{sec:restricted_connectivity}.

\subsection{Integer Factorization}

\begin{figure}[t!]
\hspace{0cm}
    \includegraphics[width=\linewidth,  trim={0.3cm 16cm 12.5cm 0},clip]{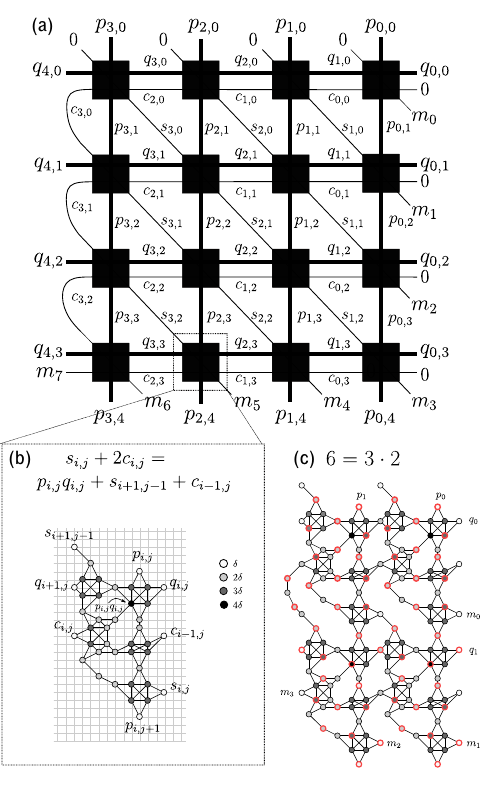}
\caption{Encoding procedure for integer factorization. (a) Graphical representation of the set of equations to be satisfied for integer factorization ($m = p \cdot q$).  The factor bits $p_i, q_i$ and binary variables $s_{i,j}, c_{i,j}$ are represented by copy lines to construct an effective square crossing lattice for the problem, with a filled square at crossings and the integer bits $m_i$ specifying some of the boundary conditions of the problem. (b) Each filled square represents a set of equality constraints between the binary variables associated with the adjacent legs. The final UDG-MWIS can be obtained by replacing each square with the factoring gadget that enforces the mathematical constraints relevant for the factoring problem. The unit-disk radius should be slightly larger than $2\sqrt{2}$ times the lattice constant. (c) An example of the UDG-MWIS representation of the factoring problem $6=3\cdot 2$. }
    \label{fig:factoring}
\end{figure}

As a final example, we now formulate the problem of decomposing an $n$-bit composite integer $m=pq$ into its prime factors $p$ and $q$ as UDG-MWIS problem. To this end, we use the binary representation for the integer $m = \sum_{i = 0}^{n - 1} 2^i m_i$, with $m_i \in \{0, 1\}$, $p = \sum_{i = 0}^{k - 1} 2^i p_i$ for the $k$-bit integer, and $q = \sum_{i = 0}^{n - k - 1} 2^i q_i$ for the $(n-k)$-bit integer. The factoring problem thus amounts to finding the unknown bits $p_i$ and $q_i$ such that 
\begin{equation}\label{eq:factor1}
    \sum_{i = 0}^{n - 1} 2^i m^i = \sum_{i=0}^{k-1}\sum_{j=0}^{n-k-1} 2^{i + j} p_i q_j.
\end{equation}
Note that $k$ is a priori unknown. However, this is not an issue, as one can consider the problem \eqref{eq:factor1} for each possible value $k = 1, 2, \cdots, n/2$, or, alternatively, consider $n$-bit representations of both $p$ and $q$.

We proceed by introducing ancillary binary variables $s_{i,j}$ and $c_{i,j}$, which may be interpreted as partial sum bits and carry bits, respectively. Using elementary algebra, the factoring problem \eqref{eq:factor1} may be expressed as a system of $k(n-k)$ coupled equations~\cite{Burges2002Factoring, Dridi2017Prime}
\begin{equation}
\label{eq:factoreq}
    s_{i,j} + 2c_{i,j} = p_i q_j + s_{i + 1, j - 1} + c_{i - 1, j}
\end{equation}
for $i=0,\dots, k-1$ and $j=0, \dots, n-k-1$, where the values $c_{-1, j}=c_{i, 0}=s_{i, -1} = 0$ are fixed and we identify $s_{0,j} = m_j$. Factoring $m$ thus reduces to finding binary values for $s_{i,j}$, $c_{i,j}$, $p_i$ and $q_j$ such that the $k(n-k)$ equations \eqref{eq:factoreq} are all satisfied. 

To embed this system of equations in a 2D plane, we use the copy gadget to copy the values of $p_i$ and $q_j$ to new ancillary variables $p_{i,j}$ and $q_{i,j}$ with the copy-constraints
\begin{align}
\label{eq:factoreq2}
p_{i,j} &= p_{i+1, j}\\
q_{i,j} &= q_{i,j+1},
\end{align}
and identify $p_i \equiv p_{i,0}$ and $q_j \equiv q_{0,j}$.
We then can write Eq.~\eqref{eq:factor1} as
\begin{equation}
\label{eq:factore3}
    s_{i,j} + 2c_{i,j} = p_{i,j}q_{i,j} + s_{i + 1, j - 1} + c_{i - 1, j}.
\end{equation}
We refer to the three equations \eqref{eq:factoreq2}-\eqref{eq:factore3} (for a given $i,j$) as the factoring constraints $F_{i,j}$. The constraints $F_{i,j}$ are manifestly local in two dimensions, in the sense that the variables $s_{i,j}$, $c_{i,j}$, $p_{i,j}$ and $q_{i,j}$ can be arranged on a square lattice such that all factoring constraints $F_{i,j}$ involve only neighboring or diagonally neighboring variables. 
A graphical representation of this is given in Fig.~\ref{fig:factoring}(a), with the box at lattice point $(i,j)$ representing the constraints $F_{i,j}$. Specifically, each line represents a binary variable, and each box represents the factoring constraints that have to be satisfied by the variables connected to it. We note that each variable enters in exactly two factoring constraints, except at the boundary.

This formulation of the factoring problem allows for a mapping to a UDG-MWIS problem. Specifically, we introduce a new factoring gadget consisting of a weighted 32-vertex unit-disk graph depicted in Fig.~\ref{fig:factoring}(b), where we identify 8 of the vertices with the 8 variables involved in the factoring constraints $F_{i,j}$. See Appendix~\ref{sec:factoringAppendix} for more details on the construction of the factoring gadget. 
In particular, we follow the design principle given in Sec.~\ref{sec:CSFMWIS}: the factoring gadget is designed such that (i) the MWIS space is degenerate, (ii) every MWIS coincides with a valid solution of $F_{i,j}$ on the vertices that represent the involved variables, and (iii) every valid solution of $F_{i,j}$ is represented by at least one MWIS. All these requirements can be checked by exhaustive search for the factoring gadget depicted in Fig.~\ref{fig:factoring}(b). Since each variable has to satisfy two factoring constraints (see Fig.~\ref{fig:factoring}(a)), we designed the factoring gadget such that this geometrical requirement can be easily met. Indeed, we can represent the full set of constraints $\{F_{i,j}|i=0,\dots, k-1;j=0,\dots,n-k-1\}$ as a unit-disk graph (of unit radius $r = 2 \sqrt{2}$ on a square lattice), by repeating the factoring gadget on a $k(n-k)$ square lattice, as depicted in Fig.~\ref{fig:factoring}(c). 
This construction therefore results in a lattice with some of the boundary conditions fixed by the values of $m_i$, such that the MWIS of the rest of the graph reveals the values of $p_i$ and $q_j$, satisfying Eq.~\eqref{eq:factor1}, thus providing the solution for the factoring problem.

\section{Numerical Simulations}
\label{sec:numerical_simulations}

\begin{figure}[t!]
 \includegraphics[width=\linewidth, trim={0 16.5cm 12.5cm 0} ]{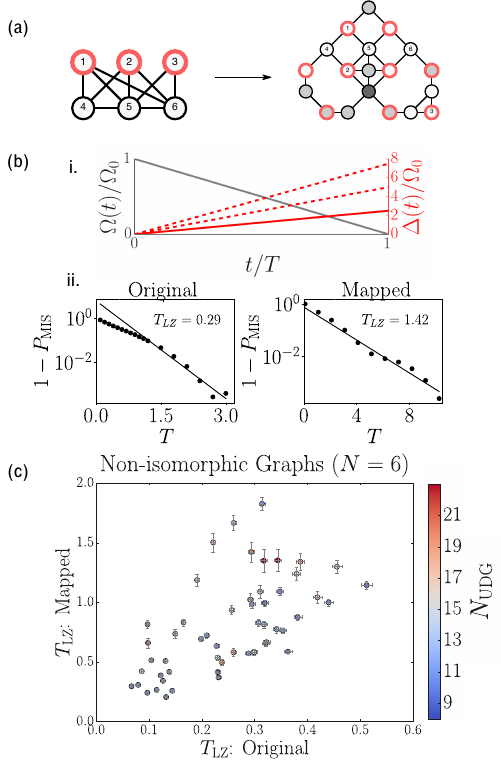}
\caption{QAA performance for MWIS on original and mapped graphs. (a) Example unit-disk mapping and MIS configuration of a 6-vertex graph. The original graph (left) is unweighted, and the mapped graph is a weighted UDG (right). The corresponding vertices are labeled with numbers. (b) i. Rabi frequency $\Omega(t)$ and detuning $\Delta(t)$ sweep used in the QAA protocol. The global detuning $\Delta(t)$ is shown in solid red, while $2\Delta(t)$ and $3\Delta(t)$ are shown in dashed lines.  ii. Computed $P_{\text{MIS}}$ for the original and mapped graphs as a function of total sweep time $T$.  The adiabatic timescale $T_{\text{LZ}}$, which is related to the minimum energy gap, is extracted from the long-time Landau-Zener fitting. (c) Scaling of $T_{\text{LZ}}$ for mapped graphs for the ensemble of unweighted non-isomorphic graphs with size $N = 6$.}
\label{fig:simulation}
\end{figure}

In previous sections, we demonstrate an encoding strategy to map a computation problem with arbitrary connectivity onto a maximum weighted independent set problem on a unit-disk graph, showing that the ground state of the mapped problem encodes the solution of the original problem. We now present numerical simulations of quantum algorithms to demonstrate the impact of the proposed mapping procedure on quantum performance. Specifically, we use the quantum adiabatic algorithm and consider the MWIS problem on an ensemble of non-isomorphic, simply connected six-node graphs, which includes three non-unit-disk graphs. For simplicity, we choose uniform weights for each graph $\Delta_v = \Delta$. Using the procedure introduced in Section \ref{sec:MWIS_section} and further simplification steps described in Appendix \ref{sec:overhead_reduction}, we map each of the original MIS problems to a UDG-MWIS problem and compare performance metrics of QAA for both problems. There are a total of 112 non-isomorphic graphs with 6 vertices. We sample 54 such graphs: due to limits of classical simulation, we evaluate only instances whose mapped graphs contain no more than 25 vertices; we also do not consider the 40 graphs that can be directly embedded as UDGs on the square lattice without any overhead.

\Fig{fig:simulation} presents the performance results for the ensemble of graphs. The QAA for the MWIS problems may be performed for a particular graph by varying $\Delta_v(t)$ and $\Omega_v(t)$ in the Rydberg Hamiltonian \eqref{eq:RydHamiltonian} \cite{Pichler2018MIS}.
Typically, the QAA is designed by initializing all qubits in the $| 0 \rangle$ state, where $\Delta(t = 0) < 0$ and $\Omega(t = 0) = 0$ (with $U > 0)$. QAA for MWIS is usually done in two or three stages \cite{Ebadi2022}. First, $\Omega(t)$ is ramped up to a non-zero value while $\Delta(t)$ is slowly tuned from negative to zero. Next, $\Omega(t)$ is ramped off while $\Delta(t)$ is slowly tuned from zero to positive. In this way, the initial state is adiabatically connected  to the ground state of the final, classical Hamiltonian whose ground state encodes the MWIS solution of the UDG. For sufficiently slow ramps, the quantum state of the system follows the instantaneous ground state of the time-dependent Hamiltonian and thus the final state corresponds to the MWIS of the graph.

In order to focus on adiabatic behavior near the gap closing point and account for the weights in the MWIS problem, we choose to initialize the state of the system starting at the end of the first stage, in the ground state of the Hamiltonian $\Omega(t = 0) = \Omega_0$ and $\Delta(t = 0) = 0$. Then, $\Omega$ is linearly ramped to zero over a time $T$ and each of the detunings are linearly ramped from zero to a final value. Note that because the weights $\Delta_v$ on each vertex may be different, the rate of change of detuning may be different on each vertex. This protocol is shown in Fig.~\ref{fig:simulation}(b).

For our numerical simulation, we work in the limit of $\Delta_0, \Omega_0 \ll U$, where the non-independent set space of the graph can be neglected (in experiments, this corresponds to the strong Rydberg blockade limit).  In this limit, we can restrict the simulation to the independent set subspace of the graph. We also ignore long-range Rydberg interactions, since the original graph does not have a geometric description.

The performance of QAA is often discussed via an analysis of the minimum spectral gap along the parameter path. However, the minimum gap alone is not sufficient to understand the time scales for adiabaticity, as the structure of matrix elements between ground and excited states can cause larger or smaller diabatic effects. Furthermore, it can be ambiguous for instances where multiple degenerate ground states exist. We therefore compare the QAA performance on the original and mapped problems by directly comparing their adiabatic time scales.
Specifically, we evaluate the adiabatic time scale by extracting a Landau-Zener time scale, $T_{\text{LZ}}$, which is the characteristic time needed to evolve the system adiabatically. $T_{\text{LZ}}$ is determined by fitting numerical results to the expected long-time behavior of the ground state probability $P_{\text{MIS}} = 1 - e^{a - T / T_{\text{LZ}}}$ at $T$ where $P_{\text{MIS}} \gg 0.99$.  This procedure is described in more detail in Ref.~\cite{Pichler2018MIS}.    

\Fig{fig:simulation}(c) presents results of this analysis, comparing the extracted Landau-Zener time scale for the original graphs in our ensemble with the Landau-Zener time scale of the corresponding mapped UDGs. The simulation results indicate that for the graphs considered, the timescale for adiabaticity of a mapped MWIS problem is correlated with that of the original problem: the correlation appears to be linear, but the limited range of data precludes a reliable fitting; The spread of the data is likely due to the specific structure of the graph, but
there is no clear dependence on the number of vertices in the mapped graph.
Unfortunately, these instances are far from the large-problem size limit and the displayed time scales give us little intuition about the asymptotic performance of the quantum algorithm for larger graphs. To have more conclusive understandings of the performance of quantum algorithms, one has to study it in Rydberg atom array experiment on larger graphs.

\section{Conclusions and Outlook}
\label{sec:conclusions}
In this work, we described an encoding strategy to map a variety of computation problems with arbitrary connectivity to maximum weighted independent set problems on unit-disk graphs, which have a  hardware-efficient implementation of quantum optimization on neutral, trapped atoms interacting via Rydberg states. The encoding incurs at most a quadratic overhead in the number of variables in the optimization problem and is thus very efficient. In addition, the mapping follows an explicit, straightforward procedure, which produces a unit disk graph that can be embedded on a square lattice with favorable conditions on the necessary unit disk radius and hence enables practical implementation with Rydberg atom arrays.

We provided three concrete examples for the problem mapping: MWIS problem on general graphs with arbitrary connectivity, the QUBO problem, and the integer factorization problem. In all examples, we show how the formation of a crossing lattice together with a few encoding gadgets enable a simple, unified approach to map a wide range of computation problems into UDG-MWIS. Numerical simulations indicate that the performance of quantum algorithms on the mapped problems is directly correlated with that of the original problems. If the linear correlation in the Landau-Zener time persists asymptotically, this suggests that any quadratic quantum speedup~\cite{Ebadi2022} in the original graph may be transferred to an equivalent speedup on the unit disk graph.

While in this work we focused on encodings into UDG-MWIS, we remark that similar strategies can also be designed for encoding computation problems into unweighted UDG-MIS problems. This is favorable for experimental implementation when local detuning capabilities are not available. The details for unweighted encoding are described in a companion paper~\cite{unweightedpaper}. There are at least several interesting future directions that deserve further explorations. First, our approach may be generalized to encode computation problems that include interactions involving three or more variables, such as higher-order unconstrained binary optimization (HUBO) problems; the NOR gate shown in Fig.~\ref{fig:CSP}(c), for example, is a constraint that involves three variables. It would also be interesting to consider encoding approaches beyond 2D geometry, for example, by generalizing the idea of a crossing lattice to a ``crossing cube" in 3D; one may design encoding methods with lower overhead in 3D or make use of the third dimension in other ways such as thinking of it as the time direction for circuit satisfiability problems. Finally, we emphasize that our encoding strategy focuses on exact (ground-state) solutions. An interesting question is to what extent such strategies can be employed for approximate optimization. It will be important to understand the effects of the encoding mappings on the performance of quantum algorithms in terms of excitations into higher-energy states.

The source code that implements the mappings in this work is available in the GitHub repository  \href{https://github.com/QuEraComputing/UnitDiskMapping.jl}{UnitDiskMapping.jl}. The (sub-)optimal configurations and weight spectrum of the maximum independent set problem instances are computed using the Julia package \href{https://github.com/QuEraComputing/GenericTensorNetworks.jl}{GenericTensorNetworks.jl}, which implements the generic tensor network algorithm~\cite{Liu2022Computing}.

\acknowledgements
We thank Leo Zhou, Soonwon Choi, Xiu-Zhe (Roger) Luo, Harry Levine, Giuliano Giudici,
Zhongda Zeng, Peter Zoller, and Nathan Gemelke for helpful discussions.
We acknowledge financial support from the DARPA ONISQ program (grant no.~W911NF2010021), the Army Research Office (grant no.~W911NF-21-1-0367), the Center for Ultracold Atoms, the National Science Foundation, the Vannevar Bush Faculty Fellowship, the U.S. Department of Energy (DE-SC0021013 and DOE Quantum Systems Accelerator Center (contract no.~7568717), the ERC Starting grant QARA (grant no.~101041435), and an ESQ Discovery Grant. Jin-Guo Liu acknowledges funding support provided by QuEra Computing Inc.~through a sponsored research program.

\appendix

\section{Overhead Reduction}
\label{sec:overhead_reduction}

In this Appendix, we provide additional strategies to the introduced mapping scheme to further reduce the overhead required for encoding the computation problems into UDG-MWIS problems.
We introduce several simplification techniques that can be easily automated to reduce the overhead of the final mapped graph, allowing us to map specific graphs with significantly less overhead.

\subsection{Crossing Lattice Reduction}

\subsubsection{Pathwidth Reduction}

\begin{figure}[b]
\centering
 \includegraphics[width = 0.9\linewidth,  trim={0 16.5cm 13.3cm 0},clip]{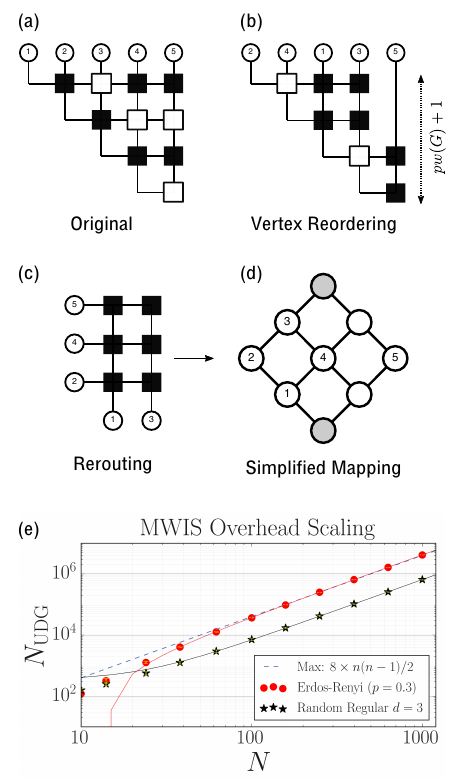}
\caption{(a)-(c) Simplification techniques to reduce the overhead. Vertex reordering can be used to reduce the depth of the crossing lattice. (d) Final mapping of the $K_{2,3}$ graph after bipartition and vertex reordering, reducing the overhead mapping from 92 nodes to 9 (see \Fig{fig:MWIS}(b)). (e) MWIS overhead scaling as a function of the number of vertices of original randomly generated graphs for chosen graph classes using a greedy pathwidth reduction algorithm.}
 \label{fig:simplificationK23}
\end{figure}

As discussed in the main text, to impose interaction constraints for arbitrary connectivity, we construct a crossing lattice.  We can reduce the depth of the crossing lattice by reordering the vertices, thus allowing the final mapping to scale with the pathwidth of the original graph. A graph $G = (V, E)$ has a pathwidth $pw(G)$ $\le k$ if and only if it has a vertex order $v_1, v_2, \cdots, v_n$ such that for any $1 \le i \le n$, there are at most $k$ vertices among $\{ v_1, \cdots, v_i \}$ that have neighbors in $\{ v_{i + 1}, \cdots, v_n \}$.  We can obtain $pw(G)$ with a path decomposition.  A path decomposition is a sequence of ``bags" ($X_1, X_2, \cdots, X_N$), where $X_i \subseteq V$ such that 
\begin{equation}
    v \in X_i, \ v \in X_k \Longrightarrow \forall j \in [i, k], v \in X_j.
\end{equation}
In other words, every vertex $v \in V$ in $G$ belongs to at least one bag and the set of bags containing $v$ forms a connected interval of the sequence $(X_1, X_2, \cdots, X_N$).  Moreover, for each edge $e \in E$, there is a bag $X_i$ that contains both endpoints. We define the width $w$ of a path decomposition as the maximum size of the bags and pathwidth $pw(G) = w - 1$.

This is advantageous because for a sparse graph, the pathwidth is usually much smaller than the number of vertices.  For example, the pathwidth of a 3-regular graph is asymptotically bounded by $n/6$, and the pathwidth of a tree graph is logarithmic in $n$.  By inspecting the appearance of the order of vertices in a bag in an optimal path decomposition, we get a good vertex reordering that reduces the size of the crossing lattice as described below.

\subsubsection{Vertex Reordering}
One can reorder the vertices in the encoding mappings to reduce the depth of the final mapped graph to the pathwidth of the graph, i.e., the size of the crossing lattice is thus $O(N*pw(G))$. More concretely, we can reduce the overhead in the mappings by minimizing the length of the copy lines and reducing the number of crossing gadgets needed. Reordering the vertices allows us to cluster crossings in the crossing lattice where the two degrees of freedom interact (such as when two vertices share an edge in the MWIS problem), and thus reduce the number of unnecessary crossings in the crossing lattice.

Graphically, as seen in Fig.~\ref{fig:simplificationK23}, for the example of the MWIS problem on the $K_{2,3}$ graph.  Fig.~\ref{fig:simplificationK23}(a) depicts the original crossing lattice discussed in Sec.~\ref{sec:crossing_lattice}. Reordering the vertices according to the path decomposition of the graph (Fig.~\ref{fig:simplificationK23}(b)) reduces the number of empty squares (crossing gadgets) from 4 to 2, thus reducing the mapping overhead even when $pw(G) + 1 = N$; in general, we will have $pw(G) + 1 < N$ for most graphs. Because the crossing lattice is a 2D mapping, we can apply the same strategy to reorder vertices along both axes: one can find a bipartition of a graph to construct a crossing lattice that minimizes the number of unnecessary crossings (or empty squares), as shown in Fig.~\ref{fig:simplificationK23}(c). Using vertex reordering, for the $K_{2,3}$ example graph, we can construct a simplified unit-disk mapping of 9 vertices (Fig.~\ref{fig:simplificationK23}(d)) whereas the direct mapping has 92 nodes.

Thus, we can generally simplify the standard mapping and reduce the overhead by restructuring the crossing lattice to reduce the length of copy lines and minimize unnecessary crossings.  The optimal vertex reordering requires computing the optimal path decomposition of a graph, which is itself an NP-hard problem. For small graphs, the optimal path decomposition can be effectively computed with the branching algorithm~\cite{Coudert2014}; for larger graphs, one can use heuristic algorithms to find good path decomposition.   This strategy allows us to achieve an overhead scaling for a chosen set of graph classes shown in Fig.~\ref{fig:simplificationK23}(e). A more detailed discussion of how vertex reordering can reduce the number of crossings is included in the companion paper~\cite{unweightedpaper}.

\subsection{Simplification Gadgets}
We can further reduce the mapping overhead from the standard encoding procedure, or any valid unit-disk mapping by introducing rewriting rules, or gadgets that maintain the integrity of the mapping, while also reducing the overhead of the graph.  Simplification gadgets are most useful for the MWIS problem, where node weights are more uniform, but simplification gadgets should preserve the weight constraints of the original problem. For example, here are some simplification rules:
\begin{center}
\includegraphics[scale=1,  trim={0 26cm 15.5cm 0},clip]{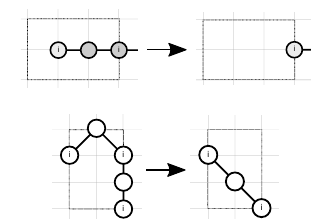}
\end{center}

\section{Defects and MWIS Guarantees}\label{sec:normalization}

As described in the main text, one need to be careful with the proper normalization of the vertex weights to ensure the ground state of the mapped problem correctly encodes the solution of original problem. In this Appendix, we provide more details on normalization for the MWIS and QUBO problem.

For the MWIS mapping shown in Fig.~\ref{fig:MWIS}, the UDG-MWIS problem is guaranteed to encode a valid solution of the original problem only if the additional weights (biases) are properly chosen. If a bias is too large, it may be energetically favorable to violate a constraint in the problem, causing the MWIS to be an invalid solution. These constraint violations, in the context of the copy gadget, are called ``defects". It is thus imperative to limit the size of the biases to guarantee that the MWIS is a valid solution.

For the constraint satisfaction problems constructed as reductions for MWIS and QUBO, there is a hierarchy of constraints. At one scale is $\delta$, which corresponds to the energy scale of the unweighted problem and at another scale is the linear ($w_i$) and quadratic ($w_{ij}$) biases that prefer certain MWIS and QUBO solutions for the weighted problems.

The safest normalization of biases is to constrain that the total weights is less than the cost of a single constraint violation. For the copy gadget, which encodes the constraint $(n_1=\overline{n}_{2})\wedge (n_2=\overline{n}_3)\wedge\cdots$, the cost of violating one constraint and adding one defect is at least $\delta$. For instance, consider a length-12 copy gadget with bias $w$

\vspace{2.5mm}
\hspace{-0.5cm}
\includegraphics[scale=0.8,  trim={0 27cm 0 0},clip]{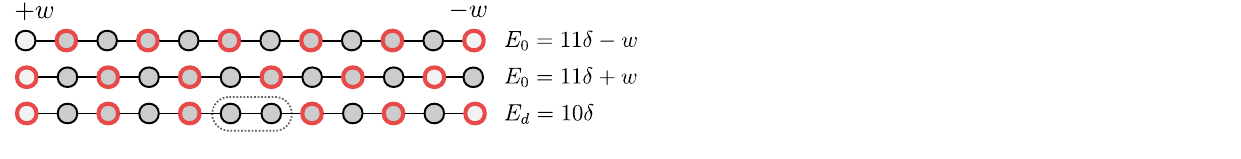}
For the configuration with a defect (the third line), the left side incorrectly represents a $1$, while the right side represents a $0$. Similarly, for the crossing gadget, removing the vertex from a clique costs an energy of at least $2\delta$, at the expense of adding two defects; one to the horizontal and one to the vertical copy gadget. Thus, the most conservative normalization of biases, which sums over all clauses, is

\begin{equation}
    \delta > \sum_{ij}|w_{ij}| + |w_i|.
\end{equation}

Unfortunately, such a normalization is too conservative. For the MWIS problem, the normalization goes as $1/N$, while, for the QUBO problem, the normalization goes as $1/N^2$. A larger bias that still guarantees a valid MWIS can be found by inspecting the structure of the copy gadget and crossing lattice.

For the MWIS problem, it is beneficial to only inspect a single copy gadget. Consider a copy gadget of length $2n$ and biases $+w$ on one end and $-w$ on the other. The valid solutions have energies $(n-1)\delta \pm w$, and the single-defect invalid solution has an energy $n\delta$. Thus, in order to guarantee a valid solution, it serves to have every bias $\delta>w_i$. In this way, the normalization must obey the constraint

\begin{equation}\label{eq:linear_guarantees}
    \delta>\text{max}_i |w_i|.
\end{equation}

For the QUBO problem, it is similarly beneficial to only inspect a single copy gadget. Single-defect solutions to the copy gadget may potentially be energetically favorable if the cost of adding a defect is outweighed by satisfying more quadratic terms. As an extreme case, consider a bit $i$ in a state $-1$, with a QUBO interaction $w_{ij}>0$ with every other qubit $j$. However, suppose the optimal state of every other qubit $j$ is $+1$ for $j<k$, and $-1$ for $j\geq k$ due to a strong linear term pinning the vertical $j$ bits. In this case, every QUBO quadratic contribution to the left of $k$ is negative while all to the right are positive, and the total contribution to the QUBO energy is small,

\includegraphics[scale = 0.9, trim={0cm 27.3cm 0 0},clip]{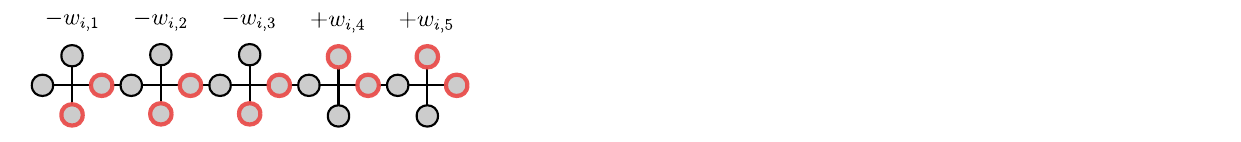}
with a weight of $w=w_i - \sum_{j<k}|w_{ij}| + \sum_{j>k}|w_{ij}|$. However, if one adds a single defect to bit $i$ at $k$, it looks like a $+1$ state to the left and $-1$ to the right, satisfying every QUBO quadratic term at the cost of adding one defect worth of energy,

\includegraphics[scale = 0.9, trim={0cm 24.5cm 0 2.5cm},clip]{inline_3.pdf}
with a weight of $w' = w_i + \sum_{j}|w_{ij}| - \delta$.
To guarantee the MWIS encodes a valid solution, the weight of the zero-defect solution must be larger $w\geq w'$. Given a QUBO contribution will always contribute a positive weight $(w>w_i)$, this guarantee is equivalent to enforcing that the defect cost is greater than the sum on all $w$ for each bit

\begin{equation}\label{eq:QUBO_guarantees}
    \delta>\text{max}_i\;\sum_j|w_{ij}|.
\end{equation}

If both the linear and quadratic constraints are satisfied by normalizing the weights correctly, the MWIS is guaranteed to be a zero-defect state and thus encode the QUBO solution.

\section{Restricted QUBO Connectivity}
\label{sec:restricted_connectivity}

\begin{figure}
    \includegraphics[width=0.95
    \linewidth, trim={0cm 15cm 13.5cm 0} ]{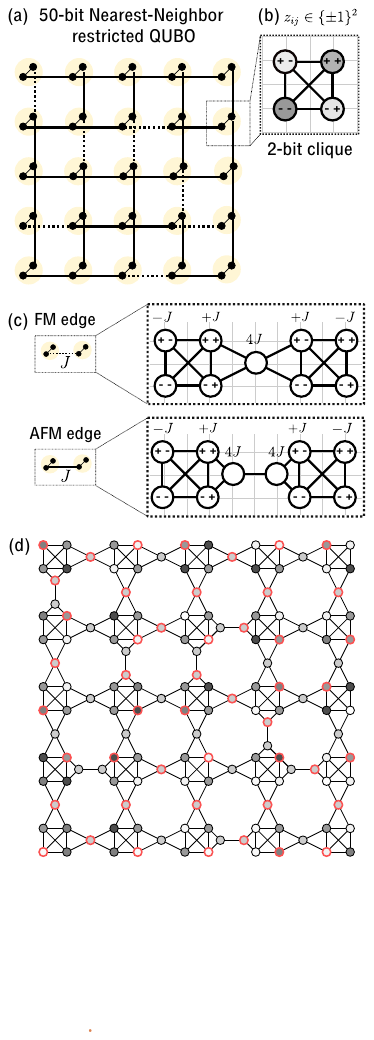}
    \caption{An example 2D restricted-connectivity graph and reduction to UDG-MWIS. (a) A particular topology of bits (vertices) and quadratic terms (edges). The original graph can be mapped into a UDG-MWIS using some gadgets. (b) Two bits can be represented by a clique of four vertices, with each state $(\pm\pm)$ represented by a single vertex of the clique. Linear and quadratic interactions are represented by biasing the weights of the clique. (c) Interactions between neighbors can be represented by adding ancilla vertices. (d) The original restricted-connectivity QUBO problem as mapped to a UDG-MWIS problem, where the ground state encodes the solution to the QUBO problem. Here, quadratic QUBO weights have been chosen to be $\pm J$. This example graph has 50 bits in the original graph and an extent of $13\times 13$ in the mapped UDG graph, which naturally fits onto today's Rydberg atom array hardware. }
    \label{fig:2d_connectivity}
\end{figure}

While it is useful to envision arbitrary-connectivity QUBO problems, its application comes at the practical cost of encoding overhead. For example, encoding a 5-bit problem takes around a $16 \times 16$ lattice, which is on the upper limit of today's Rydberg atom array system~\cite{Ebadi2022}. A more near-term solution is to specify graphs with a constrained connectivity that naturally fit more bits onto today's hardware. A natural restriction is a nearest neighbor 2D connectivity, such as an example graph shown in Fig.~\ref{fig:2d_connectivity}(a). Just like any arbitrary QUBO problem can be mapped to a UDG-MWIS problem with a quadratic overhead, a restricted 2D QUBO problem can be mapped onto UDG-MWIS with only a constant overhead.

In order to construct particular restricted connectivity QUBO problems for the particular topology of Fig.~\ref{fig:2d_connectivity}, it is instructive to introduce three new gadgets, which are extensions of the crossing gadget. The first gadget is a square clique of 4 vertices, which is similar to the 4-vertex clique of the QUBO gadget, shown in Fig.~\ref{fig:2d_connectivity}(b). The four MWIS of the clique represent four possible states of two qubits; for this mapping, we choose the top right vertex to be the $++$ state, and the bottom right vertex to be the $+-$ state, etc. In order to encode a QUBO interaction between the bits, we likewise bias the weights of each vertex as shown in Fig.~\ref{fig:2d_connectivity}(b).

An additional gadget can encode ferromagnetic $(J>0)$ or antiferromagnetic $(J<0)$ interactions between adjacent bits, as shown in Fig.~\ref{fig:2d_connectivity}(c). The independent set restriction naturally encodes $nn$-type interactions, while QUBO usually requires $ZZ$-type interactions, which can be converted back and forth using linear terms. In this way, the interaction between adjacent bits can be encoded by adding one $(J<0)$ or two $(J>0)$ ancilla vertices in between each clique within the unit-disk radius. Ultimately, the absolute value of the interaction is encoded into the weight of these interaction vertices. 

For an example of how this interaction gadget works, consider the one vertex antiferromagnetic interaction and gadget of Fig.~\ref{fig:2d_connectivity}(c). If the horizontal bit of the left clique is in the $+1$ state (e.g.,~on the right side of the clique), the ancilla vertex is blockaded from being part of the independent set, and likewise if the horizontal bit of the right clique is in the $-1$ state. Thus, the effective interaction, encoded into the condition of the ancilla vertex of weight $w_{ij}$ being included in the independent set, is $w_{ij}(1-z_i)(1+z_j)/4$. Similarly, the two-vertex ferromagnetic interaction is encoded into the weight as $w_{ij}(1-(1-z_i)(1+z_j)/4)$, as the ancilla vertex is only blockaded for one configuration instead of three. Note that the antiferromagnetic gadget has a negative sign in front of the $zz$ term, as required, and similar for the ferromagnetic gadget. After correcting the linear offsets, the biases for each vertex of the gadget are shown in Fig.~\ref{fig:2d_connectivity}(c), with $w_{ij} = J$.

Finally, one must guarantee that the ground state does in fact map to the codespace of valid solutions, with proper normalization as described in Appendix~\ref{sec:normalization}. Here, a solution is valid if each 4-vertex clique has at least one vertex in the maximum independent set. This may be guaranteed by increasing the zero-bias weight of the four-vertex clique to be much larger than any other scale. One guaranteed offset is $U = 8J$, where $J$ is the largest coupling strength between adjacent bit cliques. This condition is set because, if the weight is smaller, an independent set vertex in the clique can be replaced with two adjacent vertices of the interaction gadget, which each have a weight of $4|J|$. Thus, this bias guarantees the correct ground state. An example set of weights, which encodes a graph with random bonds $\pm J$, is shown in Fig.~\ref{fig:2d_connectivity}(d).

It should be emphasized that Fig.~\ref{fig:2d_connectivity} is just one \textit{particular} example of a local connectivity encoding for unit-disk graphs. In practice, there may be many different encodings of many different restricted-connectivity graphs. Due to the nature of Rydberg atom arrays which reconstruct the graph for each shot, these neutral-atom platforms are much more flexible in the connectivities of the problems they solve, potentially even on a shot-by-shot basis. This is in contrast with other architectures such as superconducting qubits, which have a fixed qubit connectivity and require a lengthy fabrication process to modify their topology.

Additionally, these local-connectivity graphs can encode more nonlocal problems, by increasing the ferromagnetic weight of edges such that the ground state of adjacent vertices are always the same. In this way, choosing a large ferromagnetic weight recreates the copy gadget and, by extension, may recreate the crossing lattice of the all-to-all QUBO problem shown in Fig.~\ref{fig:weighted_crossing_gadget}. Furthermore, such a local-connectivity graph may recreate other hardware’s configuration. For instance, the DWAVE Chimera graph \cite{DWAVE_topology} consists of sets of 8 bipartite connected bits in a unit cell, which are connected colinearly with adjacent unit cells. The same connectivity can be reproduced by choosing some large ferromagnetic $J$ terms appropriately on the grid of Fig.~\ref{fig:2d_connectivity}. It should be emphasized that due to the re-configurable nature of Rydberg atom arrays, it is trivial to modify the topology of the connectivity. For example, on one shot, a Rydberg atom array system could recreate the DWAVE Chimera topology, while, on the next shot, it may recreate the DWAVE Pegasus topology, and so forth.

\section{Factoring Gadget}\label{sec:factoringAppendix}

\begin{figure}[b!]
    \includegraphics[width=0.9\linewidth]{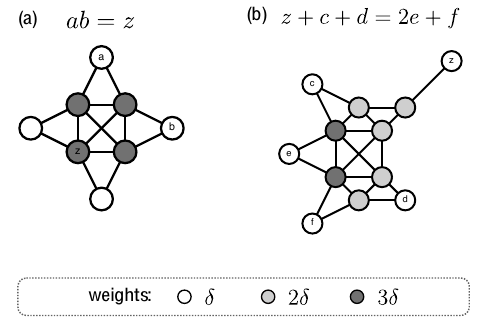}
  \caption{The two basic components of the factoring gadget. The gadget in (a) is a crossing gadget. Besides its use in routing effective variables $a$ and $b$, it also serves as a tool to access the value of the product of the variables, $z=ab$, at the indicated interior vertex. The gadget in (b) is the MWIS representation of the constraint $z+c+d=2e+f$.}
    \label{fig:math-gadgets}
\end{figure}

In this Appendix, we elaborate on the factoring gadget introduced in Fig.~\ref{fig:factoring}. The factoring gadget is designed such that the MWIS space corresponds to the satisfying assignments 
\begin{align}
    s_{i,j} + 2c_{i,j} &= p_{i,j}q_{i,j} + s_{i + 1, j - 1} + c_{i - 1, j}\label{seq:f1}\\
    q_{i+1,j}&=q_{i,j}\label{seq:f2}\\
    p_{i,j+1}&=p_{i,j}. \label{seq:f3}
\end{align}
We first focus on the constraint~\eqref{seq:f1} since the other two constraints are easy to satisfy with a combination of copy and crossing gadgets. 
To simplify notations, let us rewrite the constraint~\eqref{seq:f1} as $ab+c+d=2e+f$ between binary variables $a,b,c,d,e,f\in \{0,1\}$. We further rewrite this constraint as a conjunction of two simple constraints, namely 
\begin{align}
z&=ab\\
z+c+d&=2e+f.
\end{align}

We obtain a MWIS representation of the first constraint directly from the crossing gadget (see Fig.~\ref{fig:math-gadgets}(a)). As already discussed in the main text, the interior vertices of the crossing gadget encode the information about the variables on the boundary. Specifically, the lower left interior vertex (representing $z$) is in the MWIS if and only if both the top and the right outer vertices (representing $a$ and $b$ respectively) are also in the MWIS, thus exactly representing $ab=z$.

The MWIS representation of the second constraint is given in Fig.~\ref{fig:math-gadgets}(b). One can check by exhaustive search that the MWISs of this gadget indeed represent exactly all satisfying assignments of $z+c+d=2e+f$. To obtain the MWIS representation of $ab+c+d=2e+f$, we thus simply join the graphs in Fig.~\ref{fig:math-gadgets}(a) and (b) at the common vertex $z$. Note that the total weight of the vertex $z$ in this joint graph is the sum of its weights in each individual graph. One can easily identify this joint structure in the full factoring gadget given in Fig.~\ref{fig:factoring}(b). The remaining parts of this gadget are simply formed by combining it with copy and crossing gadgets that satisfy \eqref{seq:f2} and \eqref{seq:f3} and to route the variables to positions where they can be accessed also by neighboring factoring gadgets.

\bibliography{refs}
\end{document}